\documentclass{article}
\usepackage{dsfont}
\usepackage{graphicx}
\usepackage{amsmath}
\usepackage{subfig}
\usepackage{pgfplots}
\usepackage{multirow}
\usepackage{url}

\newcommand{\ft}[1]{\mathcal{F}(#1)}
\newcommand{\lcm}{\mathop{\text{lcm}}}
\newcommand{\partition}{\textsc{Partition}} 
\newcommand{\cpi}{\textsc{Cosine-Product-Integration}} 
\newcommand{\sat}{\textsc{Sat}}
\newcommand{\NP}{\textsc{Np}}
\renewcommand{\P}{\textsc{P}}
\newcommand{\Z}{\mathds{Z}}
\newcommand{\R}{\mathds{R}}
\newcommand{\set}[1]{\left\{#1\right\}}
\newcommand{\abs}[1]{\left|#1\right|}
\newcommand{\eps}{\varepsilon}
\newcommand{\poly}{\text{poly}}

\begin{document}

\title{Towards a Physical Oracle for the \partition~Problem using Analogue Computing}
\author{Stefan Rass\thanks{Universitaet Klagenfurt, Institute of Applied Informatics, System Security Group}
}

\maketitle

\begin{abstract}
Despite remarkable achievements in its practical tractability, the
notorious class of \NP-complete problems has been escaping all attempts to
find a worst-case polynomial time-bound solution algorithms for any of
them. The vast majority of work relies on Turing machines or equivalent
models, all of which relate to digital computing. This raises the question
of whether a computer that is \emph{(partly) non-digital} could offer a new
door towards an efficient solution. And indeed, the
\partition~problem, which is another \NP-complete sibling of the famous
Boolean satisfiability problem \sat, might be open to efficient solutions
using \emph{analogue computing}. We investigate this hypothesis here,
providing experimental evidence that
\partition, and in turn also \sat, \emph{may} become tractable on a combined digital
and analogue computing machine. This work provides mostly theoretical and
based on simulations, and as such does not exhibit a polynomial time
algorithm to solve \NP-complete problems. Instead, it is intended as a
pointer to new directions of research on special-purpose computing
architectures that may help handling the class \NP~efficiently.
\end{abstract}

\section{Introduction}
Among the vast number of problems known to computer scientists, the
structurally simplest ones seem to notoriously escape attempts towards
efficient solutions. The most prominent class of such presumably intractable
problems are \NP-complete, with the classical reference of
\cite{Garey&Johnson1979} listing hundreds of well known examples. Since all
\NP-complete problems are computationally equivalent, it suffices to solve
any problem to get the entire class tackled in one single blow.

An oracle is a complexity-theoretic concept to represent some assumption
about solvability of certain problems, so that the difficulty of one problem
can be measured relative to the difficulty of another problem (polynomial
reductions serve a similar purpose). Abstractly, an oracle is a set
(language) $A\subseteq\Sigma^*$ over some finite alphabet $\Sigma$, for which
the question ``$w\in A$?'' can be answered in constant time\footnote{The
particular style of oracle access can create subtle differences here as
discussed in \cite{Fortnow.1994}, but this is of no particular relevance
here.}. Taking $A$ as any \NP-complete problem, say assuming that we can
answer satisfiability of a Boolean formula in conjunctive normal form in
constant time, reduces the effort for any other problem in \NP~to the labour
of converting the given problem instance into an instance of $A$. In brief:
any oracle $A$ that is \NP-complete equates $\P^A=\NP^A$. It has also been
shown,
however, that there are other oracles $B$ that separate the classes as $\P^B\neq\NP^B$, 
so that the famous \P-vs-\NP~question cannot be settled by any argument that
uses oracles \cite{Baker.1975}.

Still, all classical results in this area rely on the conventional Turing
machine model or an equivalent thereof, which is intrinsically discrete (and
hence suitable for digital computing). The hypothesis put forth and
investigated in this report concerns the use of analog computing to
physically build an oracle that may be able to solve an \NP-complete problem,
thus technically equalizing $\P$ and $\NP$ under this oracle, even though
doing this outside the Turing machine model.

The problem of choice is \cpi~(problem [AN14] in \cite{Garey&Johnson1979}):

\begin{description}
  \item[Instance:] A sequence $(a_1,a_2,\ldots,a_n)$ of integers
  \item[Question:] With
  \begin{equation}\label{eqn:cosine-product}
      s_n(t):=\prod_{i=1}^n\cos(a_i\cdot t)
  \end{equation}
  does,
  \begin{equation}\label{eqn:integral}
  \int_0^{2\pi}s_n(t)dt = 0
  \end{equation}
  hold?
\end{description}
The analysis of this problem in terms of complexity and solvability is most
instantly achieved by taking the Fourier transform of $s$, which is
\begin{align}
    \ft{s_n}(\omega)=\frac{\sqrt{\pi/2}}{2^{n-1}}\sum_{M\subseteq N}&\left[\delta\left(\sum_{i\in M} a_i - \sum_{i\in N\setminus M}a_i+ \omega\right)\right.\nonumber\\
    &\quad+\left.\delta\left(\sum_{i\in M\setminus N} a_i - \sum_{i\in M}a_i + \omega\right)\right]\label{eqn:fourier-spectrum},
\end{align}
where $\delta$ is the Dirac distribution, and $N=\set{1,2,\ldots,n}$.
Equation \eqref{eqn:fourier-spectrum} is easily derived from the convolution
theorem and induction on $n$: For $n=1$, we have $\ft{\cos(a_1\cdot
t)}(\omega)=\sqrt{\frac{\pi }{2}} \delta (\omega-a_1)+\sqrt{\frac{\pi }{2}}
\delta (a_1+\omega)$, and \eqref{eqn:fourier-spectrum} holds. Now, assume
\eqref{eqn:fourier-spectrum} to hold up to $n-1$, and consider
$\ft{s_{n-1}\cdot\cos(a_n\cdot t)}=\ft{s_{n-1}}\ast\ft{\cos(a_n\cdot t)}$ by
the convolution theorem. Since $\ast$ is distributive w.r.t. addition and
associative with scalar multiplication, the induction hypothesis gives
\begin{align*}
(\ft{s_{n-2}}&\ast\ft{\cos(a_n\cdot t)})(\omega)=\\
&\frac{\sqrt{\pi/2}}{2^{n-1}}\sum_{M\subseteq N}\left[\delta\left(A + \omega\right)+\delta\left(-A + \omega\right)\right]\ast\left(\sqrt{\frac{\pi }{2}} \delta (\omega-a_1)+\sqrt{\frac{\pi }{2}}
\delta (a_1+\omega)\right),
\end{align*}
with the terms $A = \sum_{i\in M} a_i - \sum_{i\in N\setminus M}a_i$ to
abbreviate the expression. The inner convolution evaluates to
\[
\frac{\pi}{2}   \delta (a_n+A-\omega)+\frac{\pi}{2}   \delta (a_n-A+\omega)+\frac{\pi}{2}   \delta
   (-a_n+A+\omega)+\frac{\pi}{2}   \delta (a_n+A+\omega),
\]
from which \eqref{eqn:fourier-spectrum} directly follows after rearranging
terms.

The Fourier spectrum provides a variety of useful insights: first, note that
the zero-th harmonic $\delta(\omega)$ appears in the spectrum if and only if
there is some subset $M_0\subseteq N$ for which we have the identity
\begin{equation}\label{eqn:partition}
  \sum_{i\in M_0} a_i=\sum_{j\in N\setminus M_0} a_j.
\end{equation}

This is indeed a well-known \NP-complete problem, known as \partition:
\begin{description}
  \item[Instance:] A finite set $A$ and a ``size'' $s(a)\in\Z^+$ for each
      $a\in A$.
  \item[Question:] Is there a subset $A'\subseteq A$ such that
  \[
    \sum_{a\in A'} s(a)=\sum_{a\in A\setminus A'} s(a)?
  \]
\end{description}
The formulation \eqref{eqn:partition} is easily recognized as a special case
of the \partition-problem by taking the size of the element $a_i$ equal to
the (positive) integer $a_i$ itself. A slight issue remains with the signs of
the integers, since \cpi~allows negative integers in the list, but the
$\cos$-function is symmetric around 0 anyway, so the signs do not matter
here.

Going back to the Fourier spectrum, we can thus solve the \partition-problem
if we could decide whether there is a zeroth harmonic in the Fourier
spectrum. Doing this numerically would work by straightforward evaluation of
the function $s$ defined by \eqref{eqn:cosine-product}. Shannon's sampling
theorem tells us that we require at most twice as many points as the maximal
frequency in the spectrum is. Here, we get another insight from
\eqref{eqn:fourier-spectrum}, since the maximal frequency is precisely
$f_{\max} = a_1+a_2+\ldots +a_n$. Hence, we require $2\cdot f_{\max}$ points
on the curve $s(t)$ within $t\in[0,2\pi]$, to run a (fast) Fourier transform
to get the zero frequency. The latter can be done in polynomial time in the
number of points, but since this number depends on the ``magnitude'' of the
problem (i.e., the range of the numbers that describe it), the overall
running time of such a solution attempt would be pseudo-polynomial (as it is
polynomial in both, the number of points and the magnitude of the problem).
This confirms another well known statement made in \cite{Garey&Johnson1979},
who mention that \cpi~is indeed solvable in pseudo-polynomial time. As such,
it is one of the ``easiest'' among all \NP-complete problems.

So, with a digital computer being presumably bound to an undesirable runtime
(unless $\P=\NP$), our next goal is looking for what analogue computing can
do here. The crux of analog computing lies in the effect of all
(exponentially many) spectral components arising simultaneously in parallel,
rather than sequentially as would be the case on a conventional computing
architecture. After a final physical cut-off (damping) of high frequencies,
we can sample the resulting signal at a constant rate, allowing us to extract
the DC part as a reliable indicator towards a YES- or NO-answer to the
initial problem. The overall procedure to solve \partition~is the following:


\begin{description}
  \item[1. (analogue computing):] Generate the signal $s(t)$ (by analogue
      computation along a cascade of 4-quadrant multipliers; see Section
      \ref{sec:analogue-multipliers}).
  \item[2. (analogue computing):] Apply a low-pass filter with a constant
      cut-off frequency $f_0$, to enable sampling the signal efficiently
      without altering its zeroth harmonic.
  \item[3. (analogue-digital transition):] Record a sample at equidistant
      time steps $t_1,\ldots,t_m$ over at least one full period of the
      output signal $s(t)$ leaving out error terms here for simplicity. The
      result is a series of pairs
        $(t_i,y_i)$ with $y_i=s(t_i)$, for $i=1,2,\ldots$. To avoid
        unwanted phase shifts in the recorded signal, the sampling must
    start at a time when all signals are aligned, which besides time
    $T_0=0$ happens first at time $T_1=\lcm\set{a_i^{-1}|a_i\in V}$
    (leaving frequency error terms out here for simplicity). The number $m$
    of points sampled crucially influences the result of the subsequent
    Fourier analysis (via inducing noise visible the Fourier spectrum,
    exposed in simulations). That number depends on the sampling time step
    $\tau$, which must satisfy $\tau\leq 1/(2f_0)$ to enable a unique
    reconstruction of the signal (sampling theorem).
  \item[4. (digital computing):] Fourier-transform the data (say, on a
      normal computer) to distill the amplitude of the component
      $\delta(\omega)$ in the signal, i.e., the direct current (DC) term,
      in the sampled signal. Call the result $DC$.

      Then, the underlying instance of \partition, has the following
      answer, based on \eqref{eqn:fourier-spectrum}:

    \begin{equation}\label{eqn:dc-condition}
    \text{\texttt{\textbf{if}}}~ DC=0~\text{\texttt{\textbf{then}} output ``YES''
    \texttt{\textbf{else}} output ``NO''}
    \end{equation}

\end{description}

Without step 2, the procedure would require pseudo-polynomial time on a
conventional computer, but computing step 1 is as well a matter of
pseudo-polynomial complexity, since the convolution with the filter transfer
function depends on the magnitude of the \partition-instance. Therefore, the
computation of the signal $s(t)$ and its low-pass filtering are shifted to an
analogue computing circuit described next. This escape of complexity is
bought at the cost of an imperfect indicator $DC$, which, even for
YES-instances, may be nonzero, but still \emph{could} be separated from its
(expected) value for NO-instances. Thus, we will simulate a physical circuit
to get a feeling on where the boundary between YES- and NO-instances is in
terms of $DC$.

\section{Analog Multipliers and Low-Pass
Filtering}\label{sec:analogue-multipliers}

Evaluating the product $s$ is straightforward by cascading 4-quadrant
multipliers with impedance converters and amplifiers in between. The
integration could -- in theory -- also be done by an analogue circuit, but is
most easily done by a low-pass filter. The crucial point here is to escape
Shannon's sampling condition by cutting off high frequency parts from the
spectrum without touching the zeroth harmonic (i.e., the direct current part
in the signal).

Consider an ideal low-pass filter, whose Fourier spectrum is a Heaviside
function that jumps somewhere in the region $(0,\min_i a_i)$. This filter
would leave only the DC part of the signal intact, thus directly delivering
the sought output. Alas, the best that we can do in practice is using filters
that strongly damp high frequencies, but we cannot annihilate them
ultimately. However, for our purposes, this is not even necessary, since all
we need to do is damping frequencies \emph{above} a certain \emph{fixed}
limit $f_0$, since the low-pass filter leaves the DC part in any case intact,
irrespectively of its cutoff frequency. However, if we can ``disregard''
harmonics $\geq f_0$, then we can sample a fixed number of $2\cdot f_0\in
O(1)$ points on the signal to recover an approximate version thereof that
contains the sought zeroth harmonic. At the same time, the complexity of this
procedure would be no longer pseudo-polynomial, since the sampling has become
independent of the problem's magnitude (controlling the high frequency parts
of the signal).

This is the trick that we seek to put to practice in the following, however,
practical matters put us back to pseudo-polynomial complexity again, since no
physical component has unlimited and perfect behavior over the entire
bandwidth. That is, if we use an analoge multiplier that works nicely up to
frequencies of, say, $f^*$, then problem instances for which $\sum a_i > f^*$
holds will no longer be correctly evaluated on our analogue computer. Neither
can any physical low-pass filter do a perfect cut-off of high-frequencies (we
can only approximate the Heaviside function in the spectrum). Still, we can
circumvent the former issue by ``squeezing'' the frequencies into the region
less than $f^*$, simply by downscaling the problem instance's magnitude
accordingly. To see why this works, note that we can multiply
\eqref{eqn:partition} with any $\lambda>0$ without altering the problem's
answer. So, by using a sufficiently small $\lambda$, we can downscale all
integers $a_i$ up to a magnitude where $\sum_{i=1}^n\lambda\cdot a_i<f^*$, so
that our analogue multipliers can work within their admissible bandwidth
regions. This change puts the harmonics closer together so that the
separation of the DC part becomes more difficult, since the ``squeezed''
version $(\lambda\cdot a_1,\ldots, \lambda\cdot a_n)$ of the \cpi-instance
$(a_1,\ldots, a_n)$ has its closest pair of frequencies at distance $\geq
\lambda$.

The inevitably imperfect low-pass filtering in any case leaves frequencies
$>f^*$ in the spectrum, so that sampling at a rate of $2\cdot f^*$ will
create alias bands overlapping the spectral region $[0,f^*)$. In turn, we
will thus see an error in the DC part that depends on the damping of high
frequencies. This error can be made small by a proper construction of the
filter, and by allowing for a higher sampling rate, so that the alias
frequencies have small amplitudes.

\subsection{Cascading Multipliers}
\begin{figure}
  \centering
  \includegraphics[width=\textwidth]{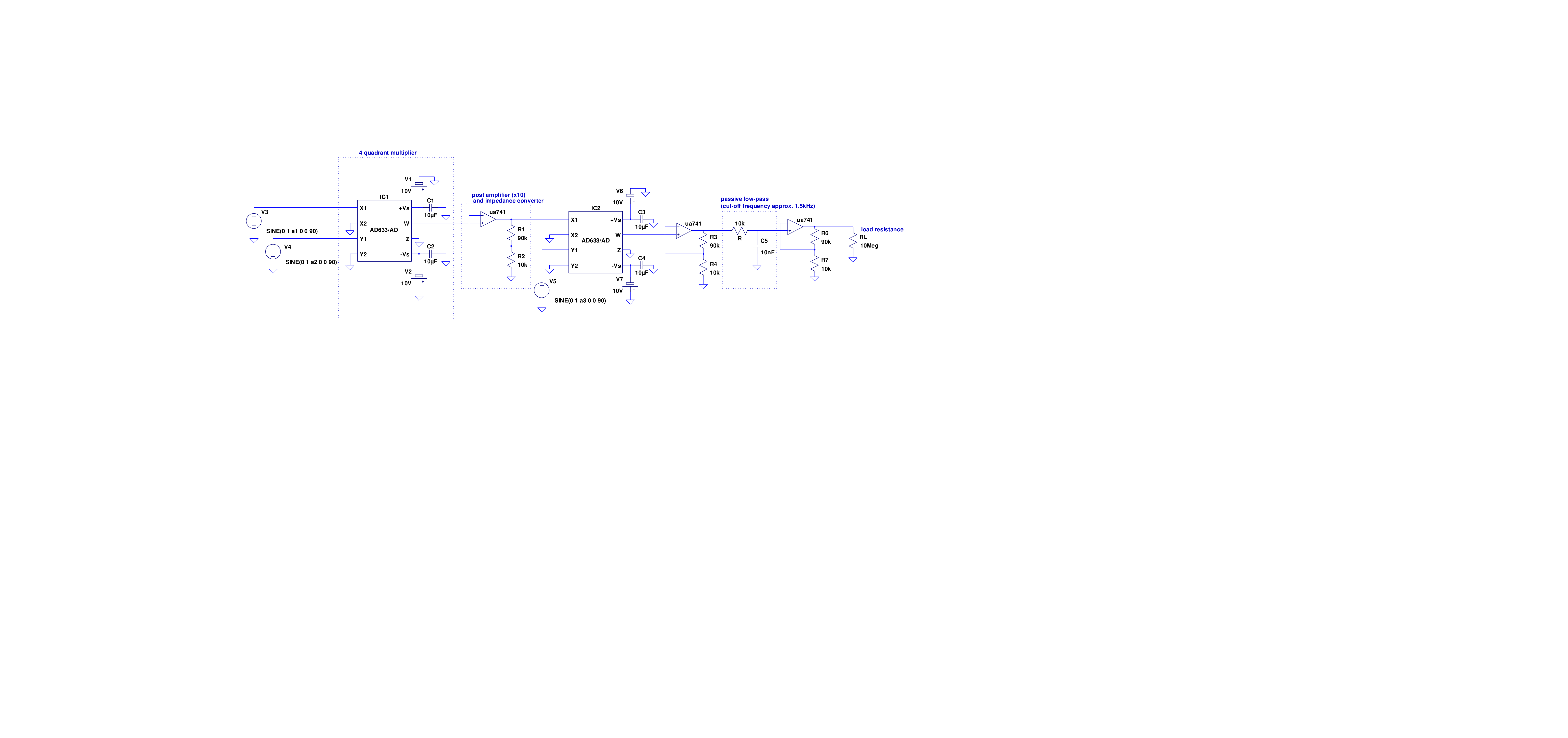}
  \caption{Example Circuit for an Example Problem (NO-)Instance $(a_1,a_2,a_3)$}\label{fig:schematic-3}
\end{figure}

Figure \ref{fig:schematic-3} displays a simple circuit that multiplies three
cosine-waves using the AD633 four quadrant multiplier by Analog Devices
(drawn in LTSpice \cite{Brocard.2013}). This component takes two signals
$x(t)=X1-X2, y(t)=Y1-Y2$ and outputs a signal $\frac{x(t)\cdot y(t)}{10}+Z$,
where $Z$ is an auxiliary additive input (the symbols in this description
correspond to the PIN configurations displayed in Figure
\ref{fig:schematic-3}). To compensate the downscaling by the factor 10 (upon
cascading the multipliers) and to have approximately ideal input and output
impedances, we add a non-inverting operational amplifier circuit in between
two multipliers. The resulting structure can then be straightforwardly
cascaded to multiply several signals, before feeding the final signal into a
passive low-pass filter and a last amplifier. The model description of the
AD633 component here directly gives the frequency window in which the
component works correctly, which ranges roughly up to $f^*=120$kHz, which is
our practical limit for $\sum_i a_i$. The same consideration must be made for
the intermediate multipliers, as frequencies accumulate along the
multiplication chain, thus increasing the likelihood to get outside the
bandwidth of the multipliers/amplifiers.

Observe that errors in the gain are of no significant relevance to our
problem, since it is a simple matter of algebra to pull out all factors
$r_i\cdot \cos(a_i\cdot t)$ in the front of the integral before doing the
analysis. So, we merely end up with a factor $r=\prod_{i=1}^n r_i$ in front
of the spectrum, but we are only interested in whether or not the integral
vanishes. Likewise, it is immediate that a multiplication of
\eqref{eqn:partition} leaves the solvability of the problem instance
unchanged. The main reason to care about amplitude errors is to avoid
steering the circuit eventually into saturation at later stages, when the
limit incurred by the supply voltage is hit. For this reason, we will work
with amplitudes of 1V per signal and 10V supply voltage, so that the
amplitude errors should not accumulate too much along the chain.

\subsection{Frequencies and Phase-Shifts}
Errors in the frequencies are trickier to handle, and induce a necessary
tolerance domain around the origin when the final decision is made. Let us
consider a distorted version $(a_i+\eps_i)_{i=1}^n$ of the original problem
instance $(a_1,\ldots,a_n)$, in which $\eps_i$ are the frequency errors.
Obviously, the answer that we get is whether or not any subset of
$\set{a_i+\eps_i: i=1,\ldots,n}$ sums up to equality of the sum over the
complement set. Put $\eps:=\sum_{i=1}^n\eps_i$. In the spectrum, we thus get
a component $\delta(\omega+\eps')$ with $\eps'\leq\eps$ for a YES-instance of
the problem, and
$\delta(\omega+\eps'')$ with $\eps''\geq \min_{i,j}(a_i+\eps_i-a_j-\eps_j)$ for a NO-instance. 
Those two are distinguishable only if $\eps$ and
$\min_{i,j}(a_i+\eps_i-a_j-\eps_j)$ are separated far enough. The difficulty
of this increases with the number of signals to be multiplied (bringing
pseudopolynomial complexity back into the game here), since squeezing the
problem instance reduces the separation of spectral frequencies and thus
calls for more accurate frequencies.

The synchronization of oscillators gains even more importance if we take
phase errors into account. Actually, the use of cosine waves in the
multiplication is crucial, as the whole idea fails if we integrate a product
of sine-waves instead (respective (counter)examples are very easy to find).
Hence, synchronization and phase shift are vital aspects in a practical
implementation, and could be done in several ways:

\begin{itemize}
  \item Synchronization of coupled analogue oscillators: this is a
      well-studied problem in the literature (see, e.g., \cite{Pikovskij.2003} to make a start), 
      but induces the issue of time until the synchronization kicks in at
  sufficient accuracy. Synchronization arguments are typically based on
  Lyapunov functions, which tell that a system will become stable (i.e.,
  synchronized), but usually remain silent on the speed at which this
  happens.
  \item Low-pass filtering of digital counters: consider a global clock
      simultaneously feeding $n$ counters, each of which triggers a
      T-Flip-Flop at its output. That is, whenever the $i$-th counter
      reaches the value $a_i$, it gets a reset signal and triggers an
      output T-Flip-Flop to produce an (approximately) rectangular pulse
      whose frequency corresponds to $a_i$. Smoothing each of these by a
      low-pass filter, and integrating them to convert the sine- into a
      cosine-waves, we end up with the desired set of synchronized signals.

      This approach, however, is not appealing for the reason of taking
      again literally pseudopolynomial time, since the counter would have
      to count up to $\lcm(a_1,\ldots,a_n)$ in order to have a fully
      synchronized set of signals for the first time.
\end{itemize}
For a lab experiment using only a few signals of low frequencies, software
like LabView can equally well do the job. For our experimental evaluation to
follow, we will put everything into a SPICE simulation, thus avoiding all
matters of imperfectness in the frequencies or phase shifts.

\section{Experimental Evaluation in LTspice}
The LTspice netlist can directly be written based on the circuit schematic in
Figure \ref{fig:schematic-3}. The analogue multiplier is Analog Devices'
AD633/AD. The operational amplifier is the (classical) ua741. We ran a
transient analysis, we chose the sampling interval, starting and stopping
time for the analysis as follows: Given frequencies $a_1, a_2, \ldots, a_n$,
the Nyquist frequency for the sampling is $f_{Nyquist}=2\cdot\sum_{i=1}^n
a_i$, so that the maximal time between taking samples is bound as
$\tau_{sampling}<1/f_{Nyquist}$. Towards synchronizing the recording of
values with the phase shift, and to allow for a burn in phase, the sampling
can start at any integer multiple of $\lcm(T_1,\ldots,T_n)$, where
$T_i=1/a_i$ is the period time of the $i$-th input signal at frequency $a_i$.
This value can be computed as
\[
\lcm(T_1,\ldots,T_n)=\lcm\left(\frac 1 {a_1},\ldots, \frac 1 {a_n}\right)=\frac
1{\gcd(a_1,\ldots,a_n)},
\]
by virtue of the general rule $\lcm(a/b, c/d)=\lcm(a,c)/\gcd(b,d)$. For our
transient analysis, we used a (maximal) time step of $2\mu$s and started
recording after $1.2$ms (the 12-fold of the $\lcm$ of periods, being 1 in all
our cases) until $3$ms. The data for the Fourier analysis is thus sampled
every $2\mu$s.

For illustration, we simulate YES and NO-instances with 3 and 4 frequencies
(for which the LTspice simulation runs reasonably fast\footnote{To fix a
convergence issue in the transient analysis, we added a shunt capacity of
$0.5$~fF by issuing \texttt{.option cshunt=2e-15} in the LTSpice netlist.
Given the frequencies that we use, the shunt capacity creates an impedance in
the giga Ohm range, so we do not need to expect too much change in the
circuit behavior. For validation of this conjecture, i.e., a verification
that the circuit behavior is not dramatically altered by the shunt
capacitance, we conducted an independent analysis using Spice Opus
\cite{Tuma.2009}, which roughly gave the same results.}).

\begin{figure}[p]
\flushleft
\subfloat[Simulated Signal $\cos(a_1t)\cos(a_2t)$ with $a_1=2\pi\cdot 20$kHz, $a_2=2\pi\cdot 30$kHz.]{
    \includegraphics[width=\textwidth,height=4cm]{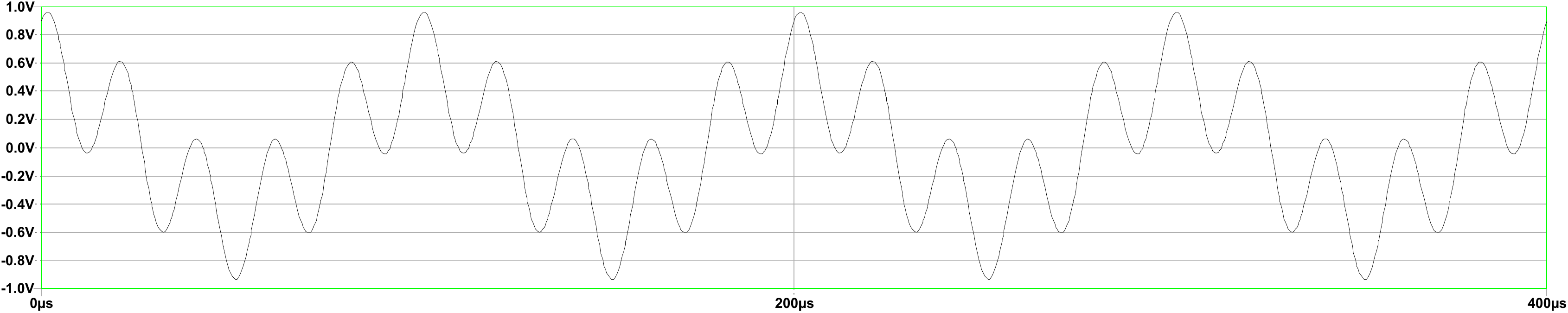}
}

\subfloat[Computed Curve $\cos(2t)\cos(3t)$]{
\includegraphics[width=\textwidth,height=4cm]{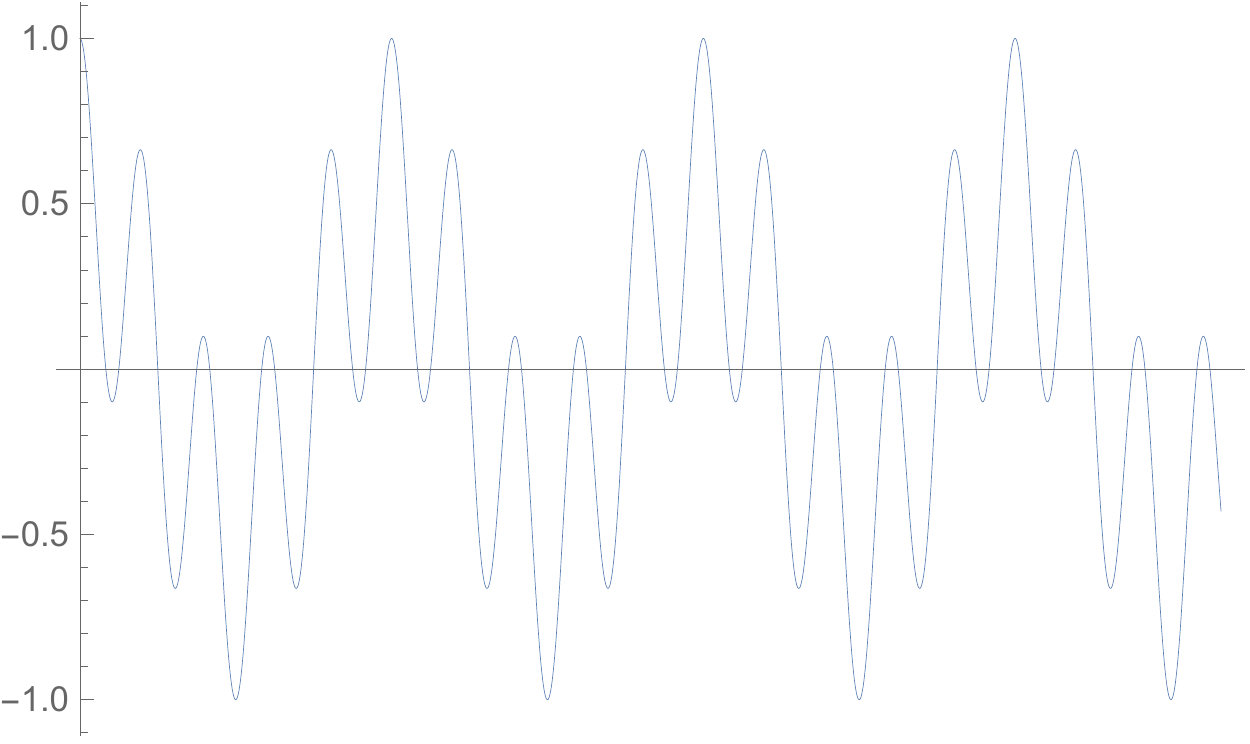}
}

\subfloat[Simulated Curve $\cos(a_1t)\cos(a_2t)\cos(a_3t)$ with $a_1=2\pi\cdot 20$kHz, $a_2=2\pi\cdot 30$kHz and $a_3=2\pi\cdot 50$kHz.]{
    \includegraphics[width=\textwidth,height=4cm]{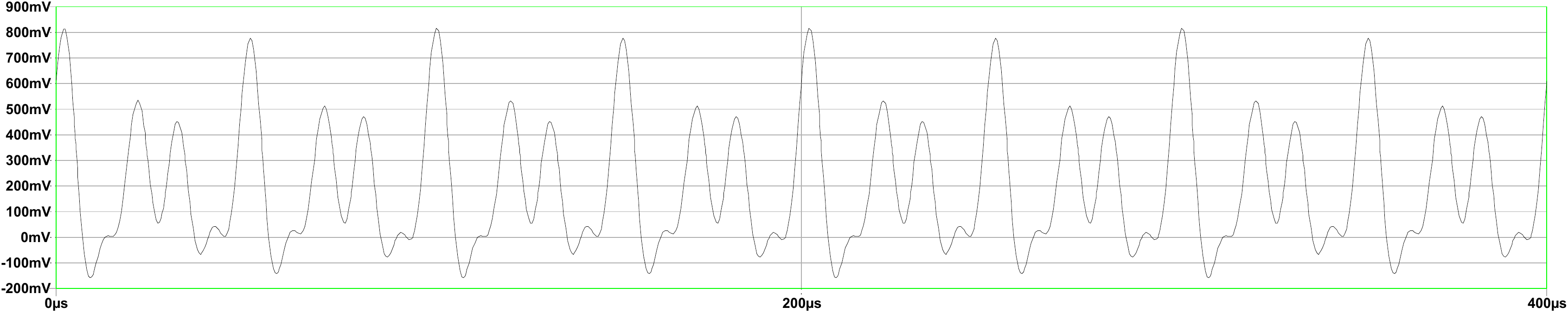}
}

\subfloat[Computed Curve $\cos(2t)\cos(3t)\cos(5t)$]{
    \includegraphics[width=\textwidth,height=4cm]{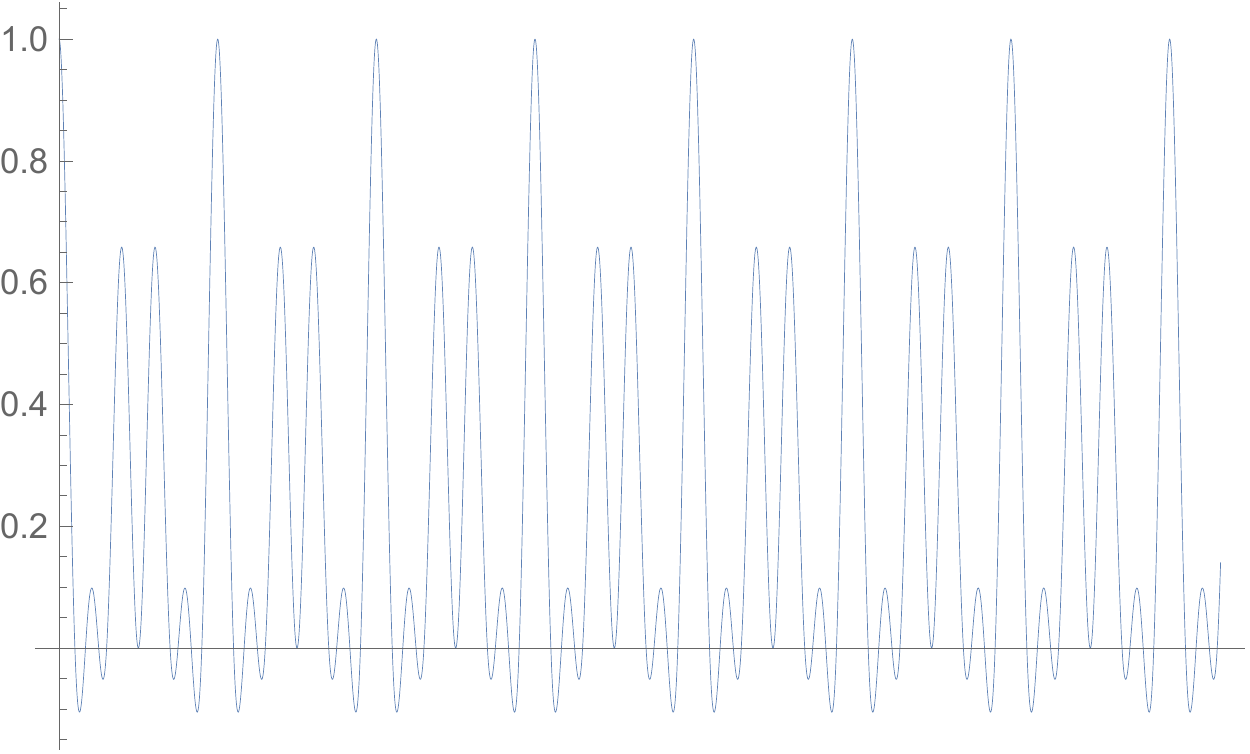}
}

\caption{Simulated vs. Computed Signals}\label{fig:waves}
\end{figure}

\begin{figure}[p]
\centering
\subfloat[Spectrum of Simulated Signal $\cos(a_1t)\cos(a_2t)$ with $a_1=2\pi\cdot 20$kHz, $a_2=2\pi\cdot 30$kHz. Theoretical Spectrum: $\ft{\cos(2t)\cos(3t)}(\omega)$ has harmonics at $\omega\in\set{-5,-1,+1,+5}$]{
    \includegraphics[width=\textwidth]{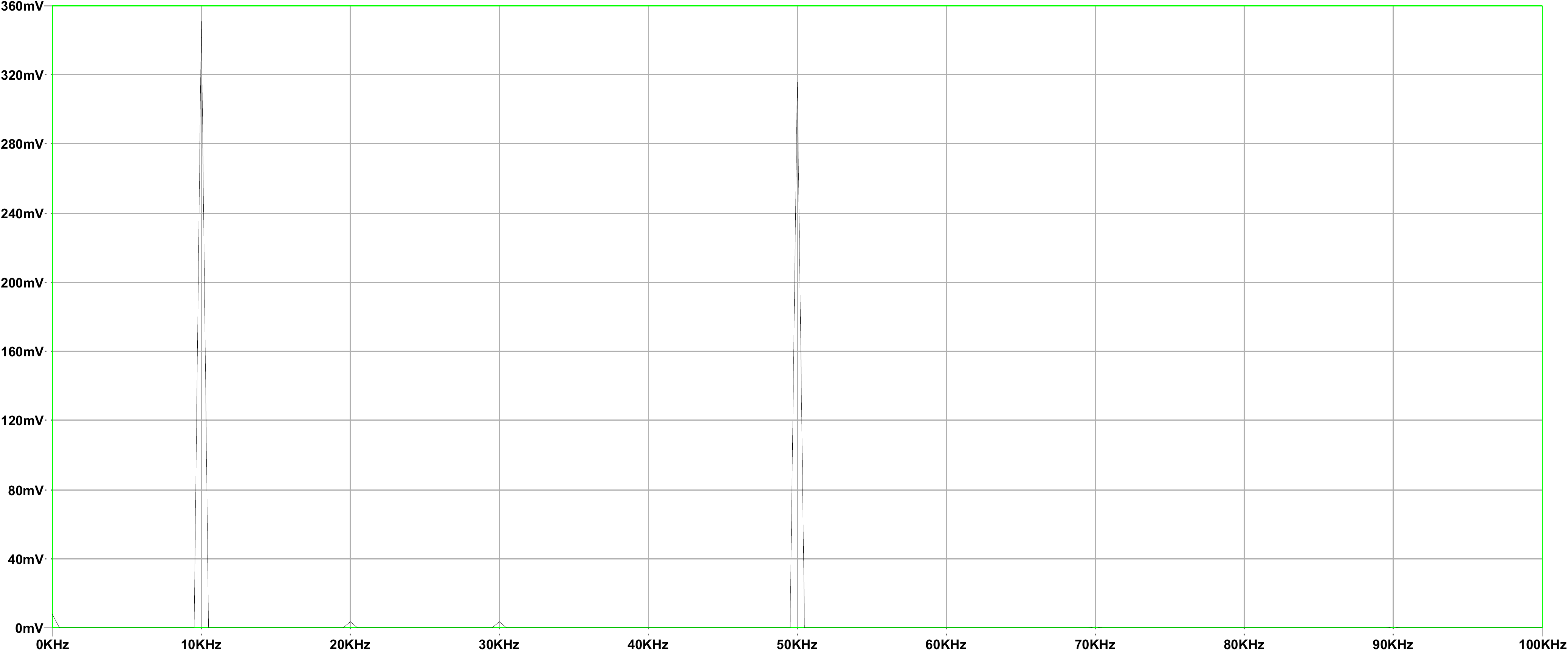}
}


\subfloat[Spectrum of Simulated Signal $\cos(a_1t)\cos(a_2t)\cos(a_3t)$ with $a_1=2\pi\cdot 20$kHz, $a_2=2\pi\cdot 30$kHz and $a_3=2\pi\cdot 50$kHz. Theoretical Spectrum: $\ft{\cos(2t)\cos(3t)\cos(5t)}(\omega)$ has harmonics at $\omega\in\set{-10,-6,-4,\textbf{0},+4,+6,+10}$]{
    \includegraphics[width=\textwidth]{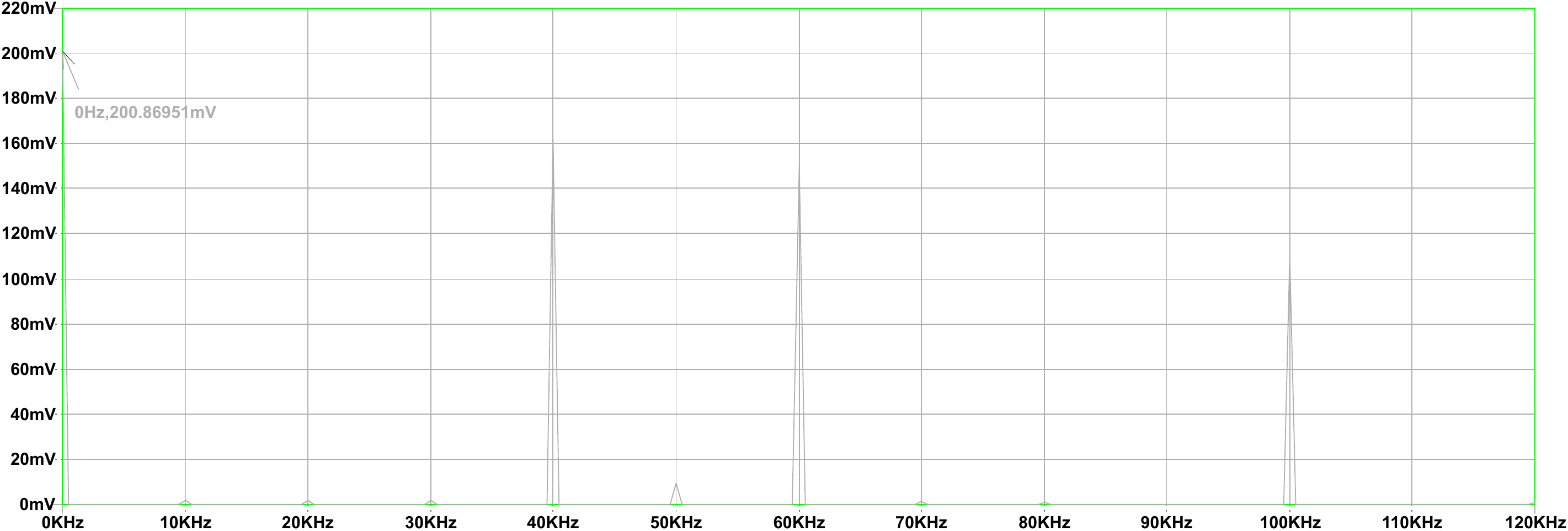}
}


\caption{Simulated vs. Computed Spectra}\label{fig:ffts}
\end{figure}

Figure \ref{fig:waves} shows the simulated vs. the ideal wave-forms, for a
rough visual validation. Figure \ref{fig:ffts} shows the respective Fourier
analyses vs. what is expected analytically from \eqref{eqn:fourier-spectrum}.
The example used for illustration is the YES-instance
$(a_1,a_2,a_3)=(3,2,5)$\footnote{The spectra obtained from the simulation are
shown with linear axis scaling; displaying the same data with logarithmic
scale (in dB) reveals a considerable lot of noise in the spectrum, which --
for the simulation -- is partly due to the discrete sampling done during the
transient analysis. Reducing the time-step in turn reduces the computed
noise, but a real circuit will necessarily show a much more complex spectrum
due to thermal and other unavoidable noise.}.

%
%
%

\begin{figure}
  \centering
  \subfloat[\ldots~for simulation]{\includegraphics[scale=0.15]{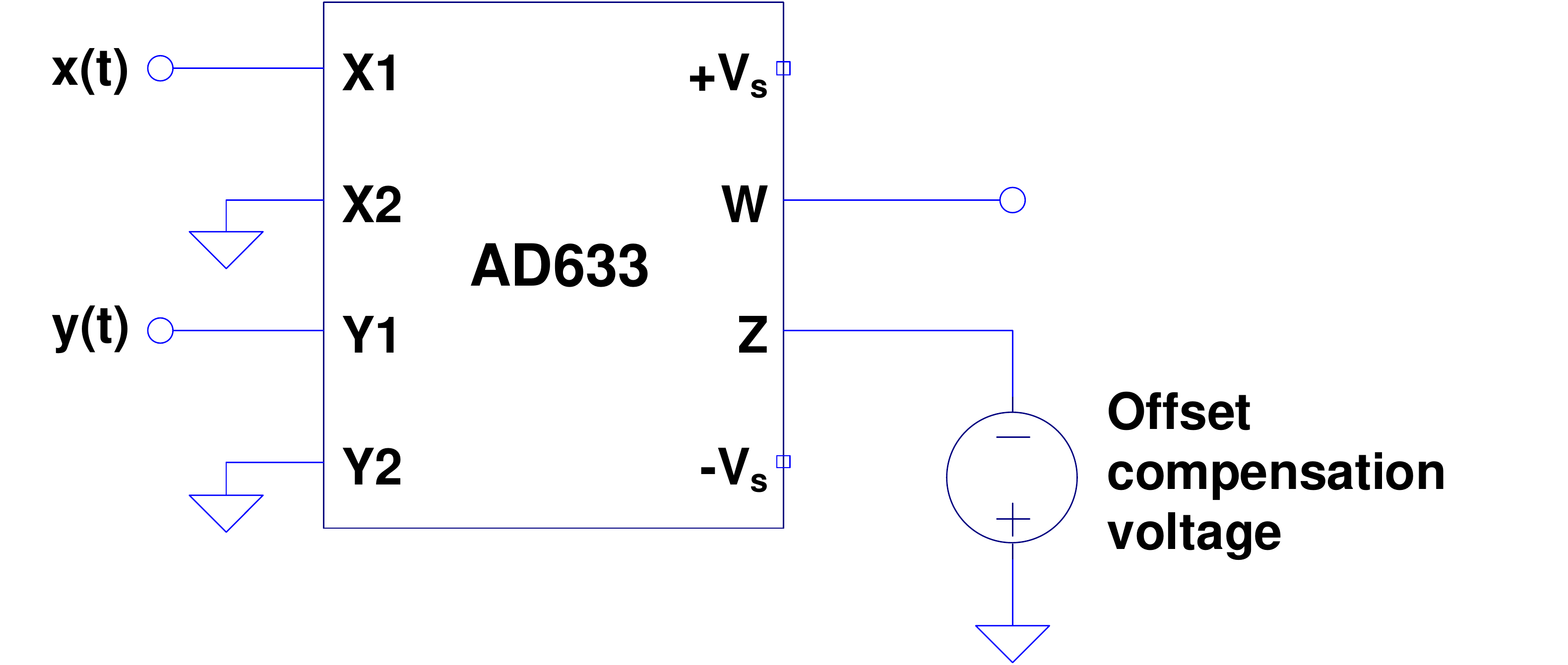}\label{fig:offset-compensation}}
  \subfloat[\ldots~for practical circuits]{\includegraphics[scale=0.15]{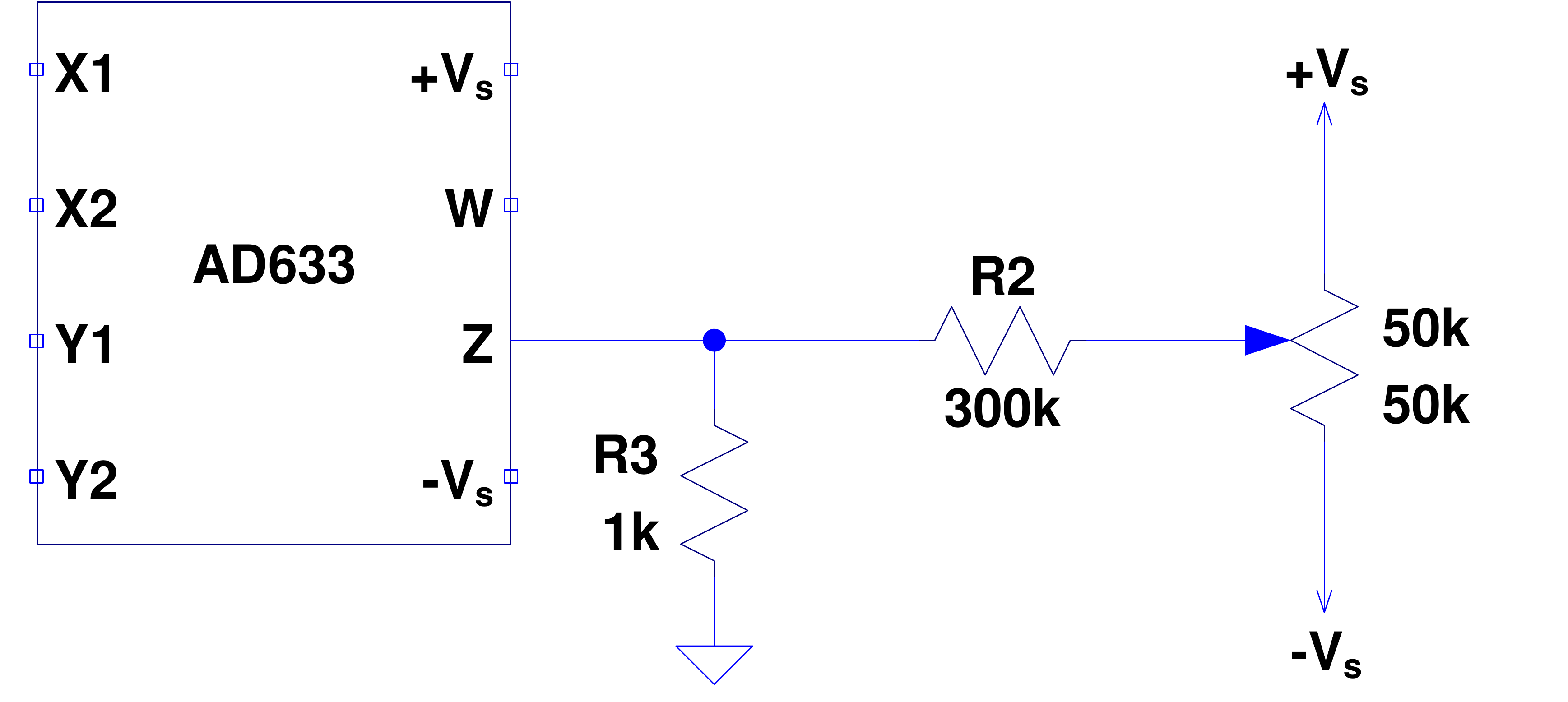}\label{fig:offset-compensation-datasheet}}
  \caption{Multiplier Offset Compensation}
\end{figure}

As the simulations indicate, the DC parts found in the Fourier analysis are
never fully vanishing for the YES-instances of the problem, although their
magnitude is clearly different between YES- and NO-instances. This effect can
be attributed to the additive offset voltages of the multiplier (typically
$\pm 5$mV up to $\pm 50$mV for the output, and typically $\pm 5$mV up to $\pm
30$mV for the inputs), as well as the operational amplifiers. Towards
compensating these, we ran a simulation using the circuit as shown in Figure
\ref{fig:schematic-3}, with subsequent Fourier analysis (\texttt{.four} Spice
directive) on the input and output nodes of each amplifier. The results found
experimentally for the NO-instance $(3,6,4)$ (corresponding to 30kHz, 60kHz
and 40kHz), was $\approx 4.22$mV at the output of the first multiplier, and
$\approx 4.31$mV at the output of the second multiplier. Since we have a
NO-instance, both should be zero, and we can use the Z-input of the AD633
component to compensate this offset by connecting it to the respective
negative potential. Figure \ref{fig:offset-compensation} shows how this is
done in our simulation\footnote{Practically, the compensation should be done
as suggested in the datasheet and displayed in Figure
\ref{fig:offset-compensation-datasheet}.}. In doing so by adding voltage
sources to the Z-pins of the two multipliers, we obtained significantly
better results. Table \ref{tbl:offset-correction} compares the results
without and with offset correction.

\begin{table}
\caption{Effects of Offset-Correction (Instances of size 3)}\label{tbl:offset-correction}
\begin{tabular}{|c|c|c|c|}
\hline
\multirow{2}{*}{Instance (kHz)} & \multirow{2}{*}{Answer} & \multicolumn{2}{c|}{$DC$ part}\tabularnewline
\cline{3-4}
 &  & without offset correction & with offset correction\tabularnewline
\hline
\hline
(30,60,40) & NO & 0.505346V & 0.0718158V\tabularnewline
\hline
(30,70,40) & YES & 2.73685V & 2.30314V\tabularnewline
\hline
\end{tabular}
\end{table}

We emphasize that the fine-tuning, i.e., offset compensation, of the
multipliers should be done by supplying NO-instances, since in these cases,
the zeroth harmonic should vanish, and whatever remains in the DC part should
be compensated. Also, the offsets were measured slightly different between
different NO-instances, which suggests that the best pick would be the
``closest'' NO-instance to the given problem $P$. The distance between $P$
and the approximate NO-instance can, for example, be measured in any norm on
$\R^n$, but a systematic search for it appears infeasible without knowing
that $P$ is actually a YES- or NO-instance (in the latter case, the sought
proximum is $P$ itself).

A feasible probabilistic approach is adding random frequency distortions to
the input signals, i.e., instantiating the circuit with frequencies
$(a_i+\eps_i)_{i=1}^n$, where $\eps_i\sim\mathcal{N}(0,\sigma^2)$ are
Gaussian error terms. Then, the random variable
$X_i:=a_i+\eps_i\sim\mathcal{N}(a_i,\sigma^2)$ is as well Gaussian. Now, let
an arbitrary subset of these be given, then \eqref{eqn:partition} fails with
probability 1 (since the event of true equality has measure zero), and the
likelihood to find a spectral line within a neighborhood $[-\Delta,+\Delta]$
of 0V (i.e., a DC-part within a given tolerance) is quantified by the
distribution of the random variable with distribution $\mathcal{N}(0,n\cdot
\sigma^2)$ to measures the event $\abs{\eps_1+\eps_2+\ldots+\eps_n}\leq
\Delta$. In choosing $\sigma^2$ properly, we can make this likelihood as
small as we desire, and get a probable NO-instance that we can use to
calibrate our circuit towards eliminating offsets.

For our experimental validation, things are simpler, as modifying the given
instances into NO-instances works efficiently. Let us give another example of
size $n=4$ to measure the offset voltages, compensate them, and then see what
we get. To this end, we added a third multiplier-and-amplifier stage to the
circuit (cf. Figure \ref{fig:schematic-3}), and did a Fourier transform at
the respective input nodes of each of the three multiplier, giving $\approx
4.5$mV for the first, $\approx 4.76$mV for the second, and $\approx 4.54$mV
for the third multiplier (note that this is indeed close to the typical
offset as told by the data sheet for the AD633 component
\cite{AnalogDevices.2017}). Table \ref{tbl:offset-correction-4} shows the
results, where a similar improvement as in the previous case can be noticed.

\begin{table}
  \centering
  \caption{Effects of Offset-Correction (Instances of size 4)}\label{tbl:offset-correction-4}
  \begin{tabular}{|c|c|c|c|}
\hline
\multirow{2}{*}{Instance (kHz)} & \multirow{2}{*}{Answer} & \multicolumn{2}{c|}{$DC$ part}\tabularnewline
\cline{3-4}
 &  & without offset correction & with offset correction\tabularnewline
\hline
\hline
(10,90,10,40) & NO & 0.514673V & 0.0675873\tabularnewline
\hline
(30,90,20,40) & YES & 0.794704 & 0.356715\tabularnewline
\hline
\end{tabular}
\end{table}

In neither experiment, we corrected any offsets induced by the non-inverting
amplifiers, but the simulations revealed that the offset contributed by these
blocks in the circuit are quite low and not of substantial magnitude (still,
they should be corrected when the circuitry is to be scaled towards large
instances). Also, some component tolerances (i.e., resistances that blur the
intended amplification factor) go into \eqref{eqn:cosine-product} as
multiplicative factors that can be pulled in front of the integral
\eqref{eqn:integral} and leave the condition unchanged.

The more important observation is the degradation of the DC-part for the
YES-instances, as is already foretold by the spectrum
\eqref{eqn:fourier-spectrum} when we consider the exponentially decaying
amplitude
\begin{equation}\label{eqn:amplitude}
\frac{\sqrt{\pi / 2}}{2^n}
\end{equation}
of all spectral components, including the DC part in particular. This decay
is necessarily exponential in the problem size, since the overall energy in
the signal gets scattered over an number of frequencies that is exponential
in $n$ (cf. Parseval's theorem).

To compensate this effect, we need to exponentially amplify the zeroth
harmonic, while applying the same exponential damping to the remaining
harmonics (otherwise, we would require an exponential lot of energy for the
amplification). The resulting filter is thus an $n$-th order active low-pass.
Such a circuit can be constructed by a chain of $n$ first order active
low-pass filters, each of which has an amplification of 2, to ultimately
cancel the denominator $2^n$ appearing in \eqref{eqn:amplitude}. Whether or
not this works practically, or whether there are other ways to make the DC
part measurable is up to experimental verification on a real circuit, and as
such beyond this report in its current form.

\paragraph{Bootstrapping:}

To reliably use condition \ref{eqn:dc-condition} in practice, it is advisable
to calibrate the analogue circuit with a set of training YES- and
NO-instances, towards identifying the voltage ranges where the respective DC
parts can be expected to be. Once this data is available, the instance under
question can be put through the circuit. Note that this method is strikingly
similar to what statisticians call \emph{bootstrapping}; cf.
\cite{Efron.1993}. It appears reasonable to use the same term to describe the
calibration and ``learning'' how the DC parts between YES- and NO-instances
are separated. The application of statistical tests is left unexplored here,
but will be part of future investigations.

%
%

\section{Conclusion}
The concept works in simulations, but these already indicate/confirm the
difficulties expected from aliasing, noise, imperfect amplitude and phase
gains, etc. It remains to experimentally verify whether careful measurements
and calibration of the circuit components can increase the magnitude
difference of the DC parts in YES- and NO-instances. Furthermore, since the
gap seems to shrink the more multipliers are in the chain, scalability of the
system is the second crucial aspect to test under lab conditions on a
physical circuit.

An independent (yet minor) theoretical caveat is the indication of the
circuit being a purely existential assertion; that is, we get only the answer
to the decision problem, but no witness of it. Finding the partition that
satisfies equation \eqref{eqn:partition} is an independent problem and can be
solved indirectly by exploiting the computational equivalence of
\partition~and Boolean satisfiability \sat. The latter is perhaps the most
important practical application of our proposed concept, so let an instance
$\psi$ of \sat~with literals $X_1,\ldots,X_n$ be given. Since \sat~and
\partition~are both \NP-complete, the given \sat-instance can be converted
into a
\partition-instance $P$ of size $m=\poly(n)$ (in polynomial time in $n$).
Since $\psi$ is satisfiable if and only if \eqref{eqn:partition} holds for
$P$, we can rule out the existence of a satisfying assignment for $\psi$ if
the DC-part coming out of our circuit is negligible. The opposite tells us
$\psi$ \emph{is} satisfiable, but we are usually interested in a satisfying
assignment too.

The latter is left untold by our analogue computer, but can be figured out by
an easy procedure as follows: put $X_1\gets 1$ (true), substitute this value
into $\psi$ and call the resulting (typically simpler) formula
$\psi\left|_{X_1=1}\right.$. Using our analogue multiplier circuit, we can
decide whether or not $\psi\left|_{X_1=1}\right.$ is satisfiable by
converting the formula into an instance of \partition. If it is, then we have
found $X_1=1$ as the first variable in the satisfying assignment. Otherwise,
we have $X_1=0$ (false) and $\psi\left|_{X_1=0}\right.$ must be satisfiable.
Let $x_1$ be the so-far correct value for $X_1$. Now, we can repeat the same
steps by guessing $X_2=1$ and checking satisfiability of
$\psi\left|_{X_1=x_1,X_2=1}\right.$, and so on, until all variables have been
determined.

In the $k$-th step for $k=1,2,\ldots,n$, the decision was done at the cost of
one conversion of $\psi\left|_{X_1=x_1,\ldots,X_k=x_k}\right.$ into a
\partition-instance, and one use of our analogue multiplier chain. The
overall effort is thus no more than polynomial in $n$, plus $n$ uses of our
analogue computer. This would -- in theory -- deliver a solution to our
\sat-instance $\psi$ in feasible time, given a constant computing time on the
analogue circuit.

As a matter of independent research interest, observe that the analogue
circuit description is constructible by a Turing machine in polynomial time
(as this merely means chaining copies of a fixed multiplier and amplifier
circuit, with changes only in the particular component properties like
voltages or resistances; disregarding offset compensations here). Still, this
is \emph{not} a uniform circuit in the usual sense of complexity theory, and
as such provides an new object to be fitted into the complexity-theoretic
landscape perhaps.

\bibliographystyle{plain}

\end{document}